\documentclass[a4paper,floatsintext,man,natbib]{apa6}
\usepackage[english]{babel}
\usepackage[utf8x]{inputenc}
\usepackage{amsmath}
\usepackage{hyperref}
\usepackage{graphicx}
\usepackage[colorinlistoftodos]{todonotes}

\usepackage{rotating}
\usepackage{rotfloat}

\title{A visualization tool for data analysis on higher education dropout: a case study at UFES}
\shorttitle{Visualization tool for data analysis on HE dropout}
\threeauthors{Pedro P. Ladeira}{Leandro M. de Lima}{Renato A. Krohling}
\threeaffiliations{
Bio-inspired Computing Lab, LABCIN \\
Federal University of Espirito Santo, UFES \\
Vitória, Brazil}
{
Bio-inspired Computing Lab, LABCIN \\
Federal University of Espirito Santo, UFES \\
Vitória, Brazil}
{
Bio-inspired Computing Lab, LABCIN \\
Federal University of Espirito Santo, UFES \\
Vitória, Brazil}

\abstract{Through the analysis of cultural, socioeconomic and academic performance aspects it is possible to map the profile of the students and their motivations to drop out. This article aims to create a computational tool for data visualization that allows drawing the profile of students to support educational institutions managers in the definition of dropout avoidance policies. We present a method to treat data collected by higher education institutions over the years, analyze them to understand the dropout and provide that information to the university and the general public. Eight questions were proposed to clarify the dropout from the Federal University of Espírito Santo, Brazil. The questions were answered through the dashboard that helps to understand the causes of dropout. It is expected that this tool can be used by others educational institutions to draw student profiles contributing to possible resolution of the problem.}

\begin{document}
\maketitle
\section{Introduction}
\label{sec:Intro}

In the last decade, the Brazilian government created several expansion programs for public higher education institutions (HEIs), which caused an increase of enrolled and dropouts students per year \citep{juniorfatores}. Student dropouts have been a challenge to the government and HEIs, and a relevant subject for academic research in several areas \citep{hoffmanninformation}. The motivations for dropouts vary from socioeconomic disparities to health problems, which makes it a complex subject and needs a study that details these motivations. Student dropout is understood, according to Brazilian Institute of Studies and Research (INEP) \citep{dosfluxo}, as the early departure, before the end of the year, series or cycle, by withdrawal (regardless of the reason), thus representing, terminative condition of failure to the objective of promoting the student to a condition higher than that of entry, concerning the expansion of knowledge, cognitive development, skills and competencies desired for the respective level of education. One of the metrics used by INEP to measure university dropout is the Accumulated Dropout Rate, also known as TDA (\textit{Taxa de Desistência Acumulada} in Portuguese), which is described in the Section~\nameref{sec:Metrics}.

According to \cite{hoffmanninformation}, student dropout can be classified in three ways: evasion of an undergraduate course, institution or system. That classifications covers changes between undergraduate course or between institutions, and academic leave of absence. INEP uses student dropout as a concept related to school or academic trajectory to describe the definitive failure in an attempt to promote the student to a condition higher than that of admission \citep{dosfluxo}. 

To investigate the profile of these individuals who drop out the higher education, it is essential to analyze the data of those students, e.g., their socio-cultural, economic and academic performance characteristics aiming to understand the main factors of these dropouts. One should also take into account which undergraduate courses these students take, the institutions where they study, the geographic regions where they reside, if they emigrated to study, if the institution has policies that seek to combat aspects that are reasons for dropouts, among other aspects. Currently, data collection is often done on a raw basis by the institutions themselves that are concerned with understanding and combating student dropout. In this work, we use as a case study the data collected in the Federal University of Espírito Santo (UFES) with the expectation of discovering a pattern of the student dropout problem, and for that purpose, an interactive graphic dashboard will be made available for managers and the general public. Using the data collected by the institution, the aim is to answer the appropriate questions regarding the profile of the students and the institution about dropout. Also, the dashboard is expected to be updated as data is collected, so that the university's current situation can be visualized and analyzed online. 

The next sections will deal with different aspects of the development of this study. First, a brief explanation of the metrics adopted in the graphs and some characteristics of the data used will be presented in Section~\nameref{sec:Metrics}. Next, the Section~\nameref{sec:Methodology} will be exposed, this section describes the dataset used in this work and the dashboard through images and explanations of the graphics. Afterward, the Section~\nameref{sec:Results} will make a connection between the graphs exposed in the dashboard section and the answers to the questions proposed in this Section~\nameref{sec:Intro}. Finally, the Section~\nameref{sec:Conclusion} discusses the relevance of this work for HEIs, proposes future works and improvements.

\subsection{Literature Revision}

The first step of the study was to describe the university dropout problem in terms of data. For this purpose, the definition of INEP \citep{dosfluxo} was used, which treats dropout as any interruption of graduation for any reason, except in the case of death, where it is considered fortuitous. INEP also defines the metric used to measure dropout: the Accumulated Dropout Rate (TDA) \citep{dosfluxo}.

The definition of student dropout in higher education can vary in different regions and countries \citep{dosfluxo,shaienkspostsecondary,spadies3,tintodefining,vossensteyn2015dropout}. It is pertinent to understand other realities to draw a parallel of similarities and differences on worldwide and Brazilian higher education. Some definitions of student dropout have been collected in the Americas, there are many points of convergence among the definitions in different regions. In Canada, the definition of a student dropping out of the university according to Statistics Canada is someone who has been studying higher education but is no longer aiming to study and has never graduated from a higher education institution or CEGEP program (\textit{Collège d'enseignement général et professionnel}) \citep{shaienkspostsecondary}. The Colombian Ministry of National Education has a program for the prevention of dropout, SPADIES (\textit{Sistema para la Prevención de la Deserción de la Educación Superior}), which defines an evaded student as a student who voluntarily or forcibly is not observed as enrolled by two or more consecutive periods, without enrolling or without being dismissed for disciplinary reasons \citep{spadies3}. In the case of Colombia, there is even a metric defined for dropout flows, which is similar to the TDA defined by INEP, that is called Cumulative Average Rate (\textit{Tasa de deserción promedio acumulada} - TDPA) \citep{spadies3}. 

In the USA, as higher education is mostly private and most of the government's concern is focused on the student's dropout of primary and secondary education, it lacks clear definitions for university dropout from institutional sources. However, the discussion about dropout exists in the USA and is widely discussed in the field of academic research. According to \cite{tintodefining} three perspectives must be taken into account in the definition of dropout: 1) of individuals who demonstrate abandonment behavior in higher education, 2) in institutions and their representatives concerned with reducing problems on their campuses and 3) in the state and nation and their respective representatives responsible for formulating policies that reduce attrition at the regional or national level. Tinto's definitions and models remain very relevant to the research field that discusses dropout from higher education, being cited in several recent works \citep{juniorfatores, hoffmanninformation, stewartfactors, meetercollege}. 

In European countries, the definition of academic success varies which different factors can be taken into account to define policies to combat dropout \citep{vossensteyn2015dropout}. The main factors are conclusion, in which students complete their studies with graduation; graduation time, in which the completion of graduation in a reasonable time is taken into account; and retention or dropout, which concerns the effort to keep students in institutions until they complete their degrees \citep{vossensteyn2015dropout}. Metrics adopted vary from country to country in Europe \citep{vossensteyn2015dropout}, even when the metric is similar, one can vary the way of calculating it, thus making a comparison between countries difficult. However, the most widely adopted metric is the completion rate \citep{vossensteyn2015dropout}, which has two forms of calculation: the cross-section method and the true-cohort method \citep{vossensteyn2015dropout}. The cross-section consists of dividing the students who graduated by the number of students entering those certain programs, that is, a few years earlier. The cross-section method is used by Austria, Belgium, the Czech Republic, Estonia, Germany, Hungary, Poland, Portugal, Slovakia, Slovenia and the United Kingdom \citep{vossensteyn2015dropout}. True-cohort calculates completion rates by tracking students who have started a program. In this case, the data used is usually old, thus not reflecting the impact of current policies to promote graduation \citep{vossensteyn2015dropout}. Countries using the true-cohort method are Denmark, Finland, France, Iceland, Italy, Netherlands, Norway, Sweden and Switzerland \citep{vossensteyn2015dropout}.

For the Brazilian context, according to \cite{hoffmanninformation} and \cite{juniorfatores}, the dropout problem was intensified with the process of increasing students admissions in Brazilian HEIs. The increase in dropout leads to a serious waste of human and material resources, both for students who drop out and for HEIs. Besides, \cite{hoffmanninformation} classify student dropout into three categories abandon of course, when the student disconnects from the course of origin without completing it (including course change); abandon of institution, when he/she leaves the HEI in which he/she is enrolled (including HEI exchange); and system abandon, when the student is permanently or temporarily absent from the academy (dropping out of higher education). The case study of \cite{hoffmanninformation} present similarities to this study, but differs in the expository sense: the graphical analysis is restricted to the article, i.e., it is used in the article as an exposition to validate an argument or theory. In contrast, an interactive dashboard has the function of exposing the data to those who use it according to the user's interest. Also, an interactive dashboard can be used for different scenarios and different universities if the input data complies with certain rules. 

According to \cite{alyahyanpredicting}, there are five aspects directly related to the academic success of a student in higher education, they are previous academic achievements; demographic data; university environment; student's psychological; academic online activity of the student. These aspects are closely related to the questions raised to be answered by the graphs. Some of these factors, such as the student's psychological and the student's academic online activity, are not captured in the data provided by UFES nor in the data provided by INEP, so they could not be analyzed in this study yet.

The panel developed in this work comes from a principle similar to that proposed by \cite{donginteractive} to track Covid-19 cases around the world, to elucidate information on a relevant issue for society, based on data that have been collected, and displayed in a clearly and interactively way to the public. The panel was created before the disease spread globally and was of great help for the population to understand the progression of the pandemic. In the case of \cite{donginteractive} the intention was to expose the number of deaths, confirmed cases and recovered cases from Covid-19 in each region of the planet. Despite different themes, the objective of the projects comes together in an attempt to elucidate statistical information on each of the subjects.  

Several contemporary data visualization tools related to higher education can be found in the literature. All the following examples deals with huge amounts of data, aims to interpret in a clear way the disorganized data and analyze information in the educational scope.

\cite{geryk2014analysis} enhanced Motion Charts tool to analyze the university retention, the result tool was named EDAIME. The EDAIME tool focuses on create animated features to successfully expose the student's behavior over time during their graduation. The EDAIME tool differs from this study in the type of analysis, while the EDAIME works in time progression of data and the animated features enhanced this type of visualization, on the other hand this study's approach do not take time in count and th graphs are static.

One featured project in the university retention matter is the FICA \citep{ferreira2020ficavis} that was created to identify and combat dropout among university students. One of the aspects of the FICA project is a data visualization tool called FICAVis. The FICAVis tool is a PowerBI based dashboard developed to highlight various aspects of the investigation on student's dropout in higher education institutions. The development process of FICAVis was made with an iteration of refining based on users evaluation and used a lot of web features, which is future desires of our present work. The FICA project seeks to identify risk situations based in several questions and act for the students in that situations. Our study tries to draw a general profile of the historical dropout student to take action through an institutional policy. Some of the risk situations proposed to be tracked by FICA project like low achievement rates and low attendance are factors that some Brazilian HEIs already keep track of, UFES for example have several study plans programs to support students with many failures in disciplines, and this is part of the dataset used in this study.

\cite{kayanda2020web} developed a Javascript web-based tool, with aid of a Python pre-processing of data to analyze dropouts in Tanzania focusing on primary and secondary schools. The visualization tool was built to fill a gap of visualizations in Tanzania about educational stats. This tool is also based in a procedure of multiples iterations and prototyping to keep refining the software with the feedback of the users, which may be implemented in the future in this work. A big difference between \cite{kayanda2020web} project and this study is that this project is focused on higher education, this change factors to be analyzed in the study and the students social context. 

\cite{deng2019performancevis} developed a web-based tool to examine students academic performance of a general chemistry course. The tool is based in D3.js and deploys multiple dashboards for different stakeholders like students, instructors and administrators. In contrast to the approach focused on a discipline, produced by Deng, this study addresses the scope of an entire university and within that scope are all disciplines in that institution.

The censuses reports released by Brazilian Ministry of Education (MEC) and INEP \citep{censoensinosuperior} are excellent examples of data visualization in the context of Brazilian higher education. The various graph formats (bar, lines, pie format) are used to illustrate relevant information from HEIs in Brazil. The topics covered range from the predominant profile of HEIs in the country to the number of students who completed the undergraduate course in the base year of the census \citep{censoensinosuperior}. 

\subsection*{Problem Description}
Dropout is defined by a student who for whatever reason decides or is forced to interrupt the undergraduate course. In contrast to the failure, it will be considered in this work also students that undergraduates in the course and students who are with the undergraduate course in progress. These three states will be the only ones possible in this study.

Possible motivations for dropout range from the individual's difficulty in staying in college due to financial, adaptation or professional issues, to the level of demand for the course and the resulting frustration. The dashboard developed in this study aims to explain the predominant motivations for the student who evades through data visualization and exploratory analysis. To this end, some questions were defined that, when answered, clarify this predominance. The questions are: 

\begin{enumerate}
	\item About the profile of the student who evades:
   	 \begin{enumerate}
        \item Does the student's income influence his/her dropout?
        \item Does the region in which the student lives affect his/her dropout?
		\item Are there courses where you can observe a large or low flow of students dropping out?
		\item What is the geographic origin of the student who evades?
        \item What is the academic performance of the student who evades? 
   	\end{enumerate}
   	\item About higher education institutions:
 \begin {enumerate}

     \item Which courses have the highest dropout rates at the institution?
         \item What is the performance rate of the highest dropout rate courses?
         \item What level of attendance of students who attend undergraduate courses with the highest dropout rates?
   	\end{enumerate}
   
\end{enumerate}

\subsection*{Motivation and Justification}

The student dropout from higher education is a social and budgetary problem that permeates the HEIs, as their function is to form economically and socially empowered individuals. It also causes a loss for the student who invests years and personal resources in pursuit of personal and professional fulfillment. When the student does not graduate as result of dropout, the HEI wastes material resources (infrastructure, laboratory materials, among others) and human resources (professors and service providers of the HEI). The main cause of dropouts, according to public and private institutions, is the lack of financial resources for the student to stay in the HEI; on the other hand, almost none of their resources is allocated to the maintenance of these individuals in the institution \citep{hoffmanninformation}. 

Aiming to tackle the causes of dropout, educational institutions have sought to understand the profile of the student who gives up on graduating from HEIs. Through socio-cultural, economic and income levels analyzes, it is possible to trace who this student is and what are the probable motivations for his/her dropout. In this work, we used as a case study the data collected over the years in one of the Brazilian HEI. Next, we process them to calculate the dropout rate and compare them to draw patterns of evasions in HEIs and make them available to the public. This will be done through a dashboard that display the graphics that will help to understand the causes of dropout. One expects that this work will serve as a tool for the institutions to identify the students' issues, helping in a possible resolution of the problem. 

Other motivation is to get a better understanding of the role of data analysis in the educational scope. Highlight the importance of good analysis and improve the quality of data can improve various aspects of the educational system. \cite{renz2020prerequisites} points that the low understanding of data is a barrier for the artificial intelligence in educational systems.

In Brazil, specially for public HEIs, their efficient management is of high importance to the population and must be object of constant improvement. Therefore, such a tool for the analysis and management of public HEIs can provide greater transparency and dissemination of accessible information.

\section*{Metrics} \label{sec:Metrics}

Since we aim to expose information understandably, one should choose the metrics used with discretion and contextualize them, so as not to cause misinterpretations or ambiguities. The metrics also provide reference points that measure the magnitude of certain situations and can expose positive or negative scenarios depending on the context. For instance, there are the mortality metric for disease impact analysis, income \textit{per capita} to analyze income distribution in a region, population literacy as a basic education metric, among others. However, the metrics themselves are not necessarily complete and out of context they can confuse and generate misinterpretations, but they are useful indicators if they are explained clearly. 

In this study, several metrics were used, but two should be highlighted because they are directly related to Brazilian higher education: the TDA \citep{dosfluxo}, created by INEP to measure dropout rate; and the UFES Student Performance Coefficient (or CR as \textit{Coeficiente de Rendimento} in Portuguese), a metric used to quantify the performance of each student during their undergraduate studies \citep{regimentoufes}. 

In case of student dropout, it is important to define a general metric to measure the dropout rates, so that there is a primary notion of the situation. Then through the analysis of the data, look for the possible causes of these indicators, whether they are high or low. \cite{silvaevasao} defines a dropout metric based on the flow of incoming and graduated students that has been widely used in other studies. Subsequently, INEP uses TDA in studies accompanying several classes, including the datasets used in this article. As it is an official metric of the Brazilian government, it was adopted and might be used in other studies for Brazilian HEIs. 

Academic performance is usually measured in Brazil through weighted averages of grades using weights that usually represent the workload for each discipline. Several factors may differ between different HEIs, these factors include the way of calculating the performance ratio, name of performance factor, curriculum of certain undergraduate courses, grades for approval in subjects, among others. UFES itself has three types of performance coefficients, two related to students and one related to courses. In this study, the student's performance coefficient was chosen, the most standard at the university, since everyone involved at some point uses this index. 

As metrics of academic performance known worldwide, we can highlight the GPA, related to the performance of students in American high school. It is used as a method of admission to US HEIs. It consists of a weighted average according to the course load, ranging from 0 to 4 and displayed with A, B, C, D or F. In universities, the forms of assessment can vary, with universities that do not even use numerical scores, using only success or failure in the discipline. 

\subsection*{Accumulated Withdrawal Rate}

The Accumulated Withdrawal Rate, or TDA, is one of the metrics used by INEP to calculate the dropout rate in each class entering a certain HEI. According to INEP, TDA can be described by an equation that involves dropout and/or transfer students, the total number of incoming students and the deceased students of a certain class \citep{dosfluxo}. First, a class is determined as those entering a course in a given year, the $ TDA $ will be a reference related to a year $ T $ of the progression of this class. The class will progress over the years with students dropping out, transferred, graduated and, in some cases, deceased. The $ TDA $ numerator ($ N_T $) can be defined by the sum of students who have withdrawn or transferred from a given class progressively, as described by:

\begin{equation}
\label{eqnumtda}
N_{T} =   \displaystyle\sum_{i=1}^{T} Des_i + \displaystyle\sum_{i=1}^{T} Transf_i
\end{equation}

\noindent where $ Des_i $ is the number of withdrawn students in the class progression year $i$, $ Transf_i $ is the number of transferred students in the class progression year $i$.

The denominator ($D_T$) is defined by the number of students who joined that class ($ Ing $) subtracted by the number of deceased students ($ Fal_i $) in the class progression $i$, as described by:

\begin{equation}
\label{eqdentda}
D_{T} =  Ing - \displaystyle\sum_{i=1}^{T} Fal_i  
\end{equation}

Finally, the $ TDA $ for that year $ T $ is defined by the percentage relative to this division as described by:

\begin{equation}
\label{eqtda}
TDA_{T} = \frac{N_{T}}{D_{T}} \times 100 
\end{equation}

\subsection*{Performance Coefficient at UFES}

According to the UFES General Regulations: \textit{``At the end of each academic term, as well as the entire undergraduate or graduate course, each student will be assigned a performance coefficient (CR) to be expressed by the quotient between the total points accumulated and total credits requested.''} \citep{regimentoufes}.

The CR is calculated by:
\begin{equation}
\label{eqcr}
CR =  \frac{\displaystyle\sum_{j}^{n}N_{j}\times C_{j}}{\displaystyle\sum_{j}^{n}C_{j}}
\end{equation}

\noindent where $ N_j $ is a grade in a discipline $ j $, $ C_j $ is the credit load of the discipline $ j $ and $ n $ is the total number of courses taken by the student. This is the sum of the multiplication of grades by the course workload, divided by the sum of the workloads already taken at the undergraduate studies.

\section*{Methodology}
\label{sec:Methodology}
The analysis was divided into 3 sections. First, section ``General'', which deals with general numbers related to the situation of students and characteristics related to dropout in the university. ``Analysis of Dropout Students'', which refers to the profile of the dropout student. Finally, ``Disciplines'' which analyzes failure rates by course and disciplines. In the case of this panel, all analyzes are focused on UFES and its students.

The end product is a dashboard with three tabs related to the sections mentioned, the ``General'' and ``Analysis of Dropout Students'' tabs have sub-tabs. In the case of the ``General'' tab, there is a sub-tab referring to the situation of students concerning the university in the quantitative analysis and another sub-tab that explains data from INEP with TDA at UFES and the comparison with national averages.

\subsection*{Dataset Used}

It is essential in data visualization to know the dataset available for the analysis, its characteristics and content determine what questions can be answered and how the data must be pre-processed so that the desired visualization be possible. So, there are two sets of data to be analyzed, one nationwide and the other from the case study at UFES. The one with national characteristics is in the public domain and can be used for analysis by any institution recognized by INEP that had classes entering in 2010. The one related to UFES is of strict use and must be requested from the university authorities. 

The data used in the analysis are provided by the institutions in tabular data format, INEP, for example, makes its data available in XLSX format, while UFES made its data available in CSV format. Pandas, a Python library, has a series of functions that transform tabular data, as well as the available spreadsheets, into Dataframes, which use the functions \textit{pandas.read\_excel} and \textit{pandas.read\_csv} \citep{mckinneypython} to read the data provided by INEP and UFES, respectively. 

Sometimes, the way data is stored in files or databases is not the right format for a particular task \citep{mckinneypython}. This is the case of UFES, after all, it is somewhat disorganized for this analysis because each line represents a discipline done by a student in a certain period and year. It is important to note that part of the information presented in the UFES dataset is anonymized. The UFES case requires data filtering, as most data needs to be pre-processed. This does not happen in the INEP data load, the pre-processing problems are limited to the selection of relevant data for analysis and header processing.  

\subsection{Dashboard}

The panel was divided into three major sections: ``General'', ``Analysis of dropout students'' and ``Disciplines''. Each section has a tab that may or may not have subdivisions. This structure allows, according to the demand, new sections and subdivisions to be added.

For this panel to become viable, it was necessary to make decisions about which programming languages and libraries would be applied. The Python language was used and the Pandas and Bokeh libraries were also predominantly used. The Python language has proven to be extremely useful in data processing and graphics display. Using the Pandas library, it was possible to handle data simply and intuitively, while the Bokeh library presented a lot of versatility for data visualization.

\subsubsection*{Panorama of UFES Students}

This section presents an overview of the university's students as a whole considering data such as number of students in each situation (Graduated, No Dropout or Dropout) at the university, number of students in each situation in each course and dropout rate per course. The failure rate used in this part of the study is the percentage of students who dropped out of graduation or were forced to drop out (metric used by INEP).
 
\begin{sidewaysfigure}[p]
    \includegraphics[width=1.0\textwidth,height=0.9\textheight,keepaspectratio]{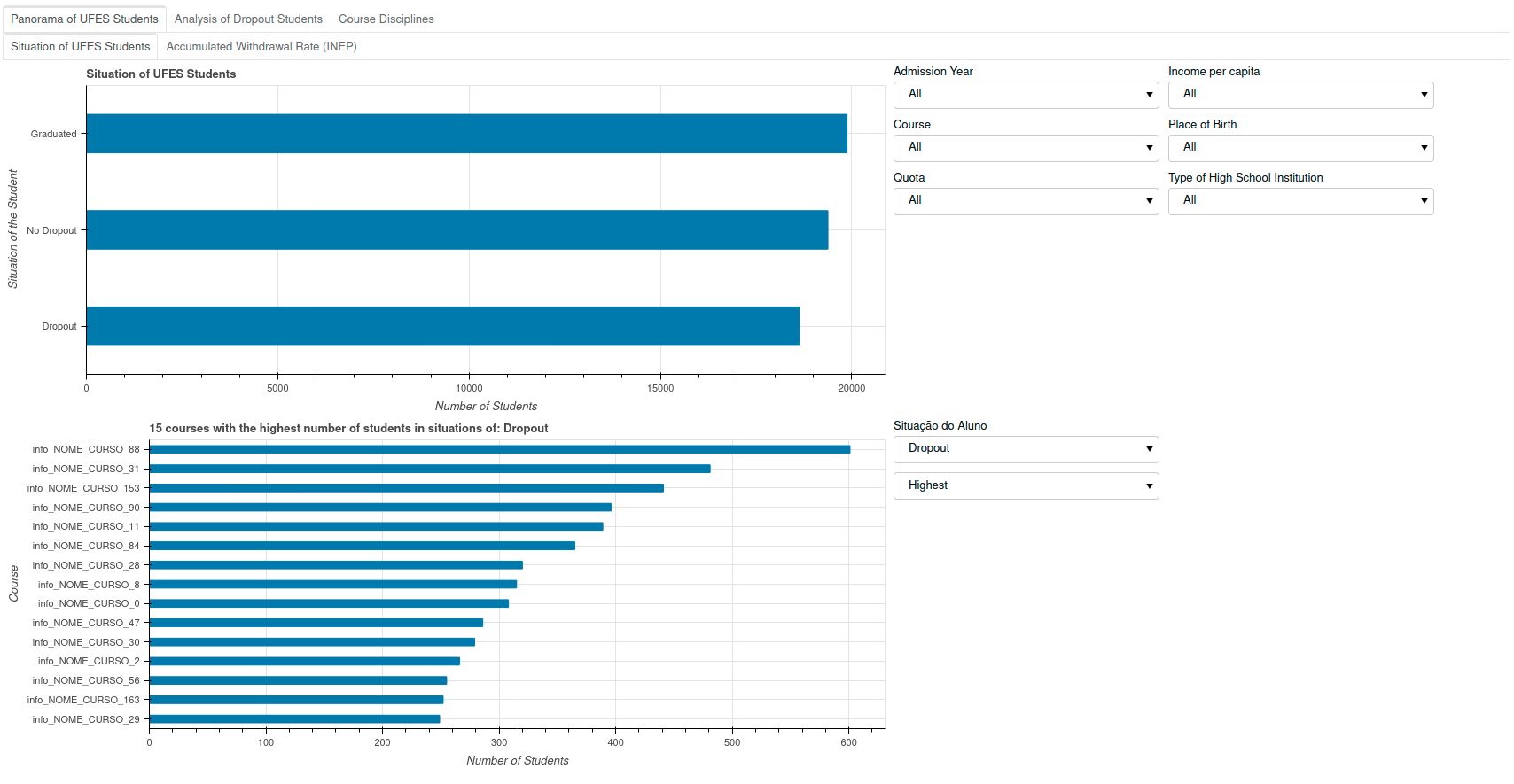}
  \caption{Graphs on the ``General'' tab, sub-tab ``Situation of UFES Students''.}
   \label{figura1}
\end{sidewaysfigure}

The composition of this tab is made up of two sub-tabs, one focusing on the situation of UFES students regarding enrollment and the other focusing on an analysis of the TDA of UFES courses. Figure \ref{figura1} shows the General tab that seeks to answer questions 1.c) ``Are there courses where you can see a large or low flow of students dropping out?'' and 2.a) ``Which courses have the highest dropout rates at the institution?''. This tab also provides a panorama of dropout mapped by courses. 

\subsubsection*{Situation of UFES Students}
\begin{sidewaysfigure}[p]
    \includegraphics[width=1.0\textwidth,height=0.9\textheight,keepaspectratio]{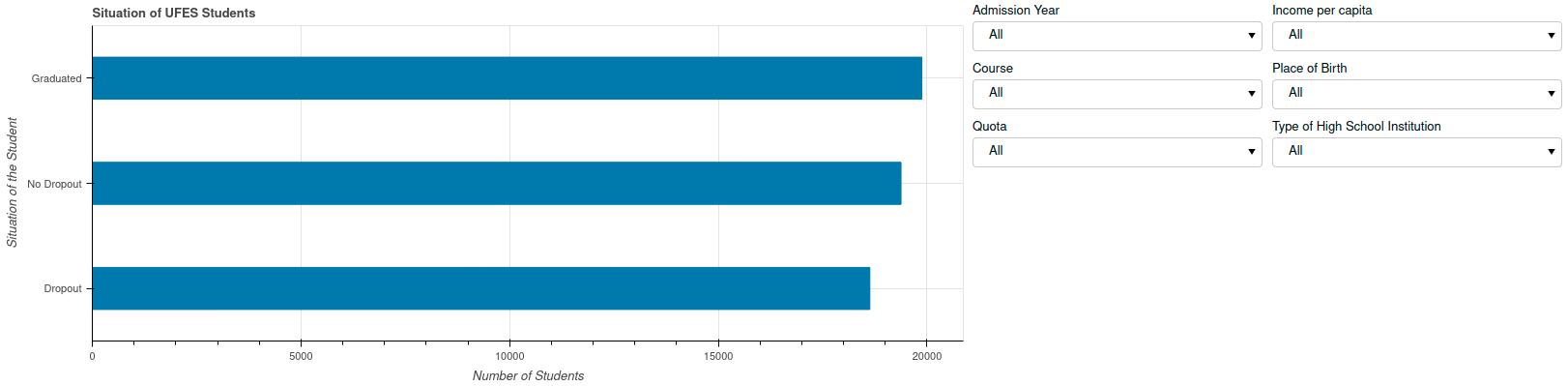}
  \caption{Graph ``Situation of UFES Students''. This graph shows the number of students in each of the three defined situations: Graduated, No Dropout or Dropout. It can be filtered by six factors that can be chosen by the user.}
  \label{figura2}
\end{sidewaysfigure}

The first graph of the general tab and the ``Situation of UFES Students'' sub-tab has the same name as the sub-tab, is shown in Figure \ref{figura2} and is positioned at the top of the sub-tab. In it there is an overview through a horizontal bar graph, of how many students at UFES have graduated, are attending or have given up or needed to interrupt the course. It is possible through multiple filters in this graph to limit the data for a course of interest, year of student entry, income \textit{per capita} of family, place of birth, type of university quota (in Brazilian public universities there is a quota system, which reserves a percentage of admissions for students of certain ethnicities and/or family income) and/or type of institution that attended high school.

\begin{sidewaysfigure}[p]
    \includegraphics[width=1.0\textwidth,height=0.8\textheight,keepaspectratio]{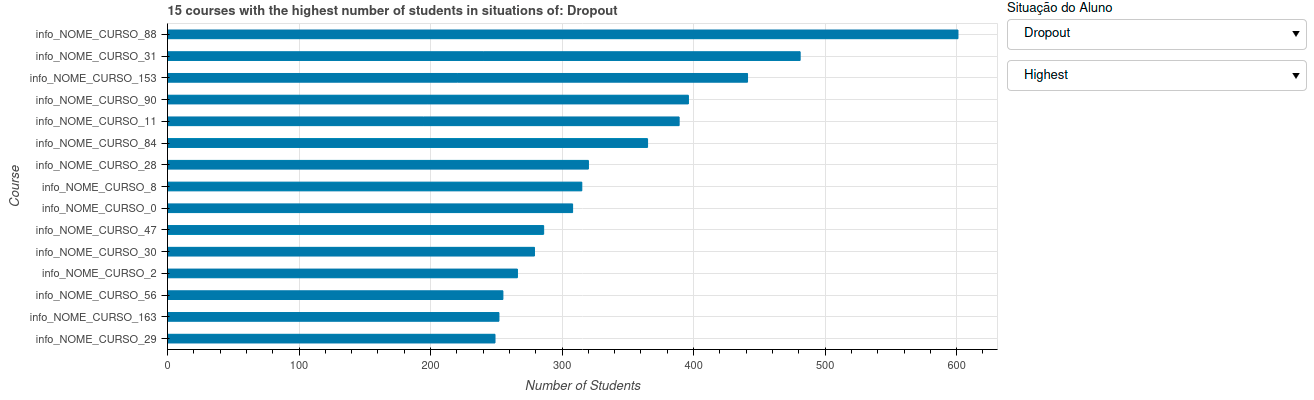}
  \caption{Graph showing 15 courses with more or less students in a given situation. Possible situations are Graduated, No Dropout or Dropout}
  \label{figura3}
\end{sidewaysfigure}

The second graph of the ``Situation of UFES Students'' sub-tab shown in Figure \ref{figura3} and located at the bottom of the sub-tab, indicates the courses that have more or less students in a given academic situation. This graph is also represented by horizontal bars and serves as a support to the first graph. The focus is on analyzing each academic situation and highlight the courses that have more and fewer students in each situation. Several factors can lead a course with a greater or lesser number of students in a given situation, such as: year of foundation of the course, annual contingent of new entrants, among others. This chart needs context for the analysis to be done.

\subsubsection*{Accumulated Withdrawal Rate (INEP)}

\begin{sidewaysfigure}[p]

   \includegraphics[width=1.0\textwidth,height=0.9\textheight,keepaspectratio]{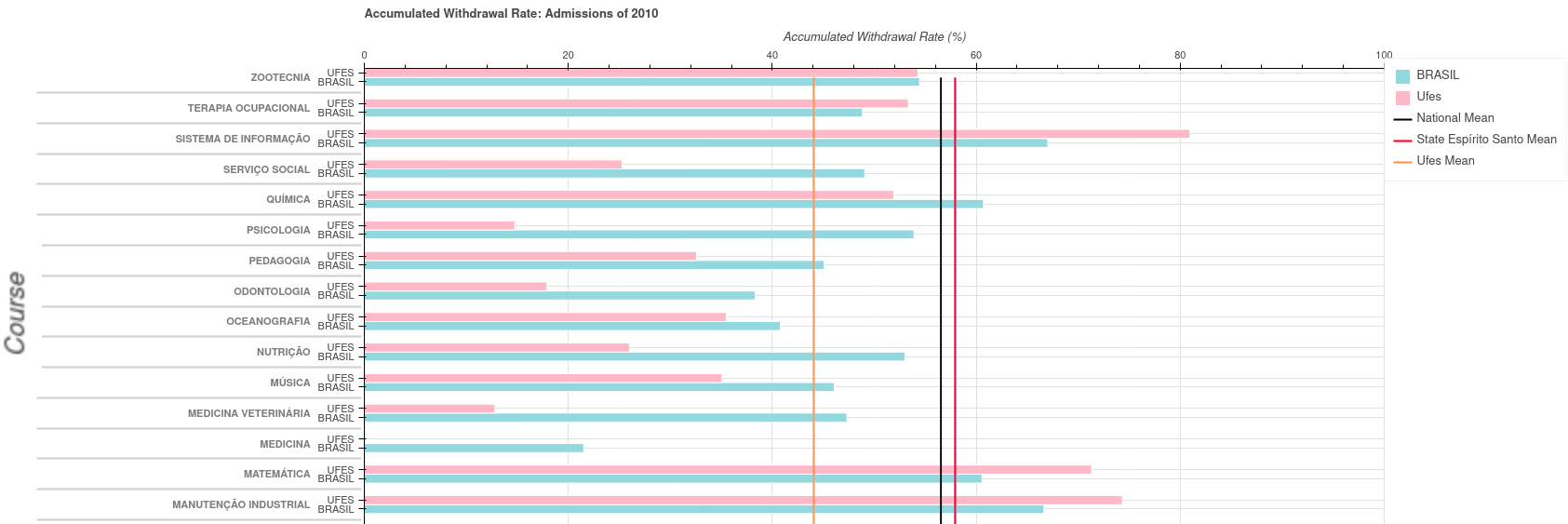}
  \caption{INEP's TDA histogram located in the ``Accumulated Withdrawal Rate'' subtab.}
  \label{figura4}
\end{sidewaysfigure}

The only graph that constitutes this sub-tab is a histogram that focuses on the TDA index shown in Figure \ref{figura4}. TDA is a parameter applied to each class in the university and is the percentage of dropout students. In the graph, the TDA is considered at the end of the class follow-up by INEP, i.e., 2016 or until the end of the class.

This graph consists of two nested bars for each UFES course. The pink bars refer to the average TDA for a given UFES course while the blue bars represent the Brazilian average TDA for all classes in that course in Brazil. In addition to the nested bars, three reference lines represent the overall averages for TDA in Brazil, the state of Espírito Santo and UFES. 

\subsubsection*{Analysis of Dropout Students}

The ``Analysis of Dropout Students'' section is the largest section of the panel in terms of graphics. This tab was divided into two sub-tabs, one referring to students' academic performance and the other referring to geographic and socio-economic indexes. The objective is to outline the recurrence profile among dropout students at UFES. This section only considers students who have permanently interrupted the course for whatever reasons, such as reopening the course, simply dropping out or having difficulty staying in college. Deaths are not considered in this analysis.

\subsubsection*{Academic Performance}

\begin{sidewaysfigure}[p]
  \includegraphics[width=1.0\textwidth,height=0.9\textheight,keepaspectratio]{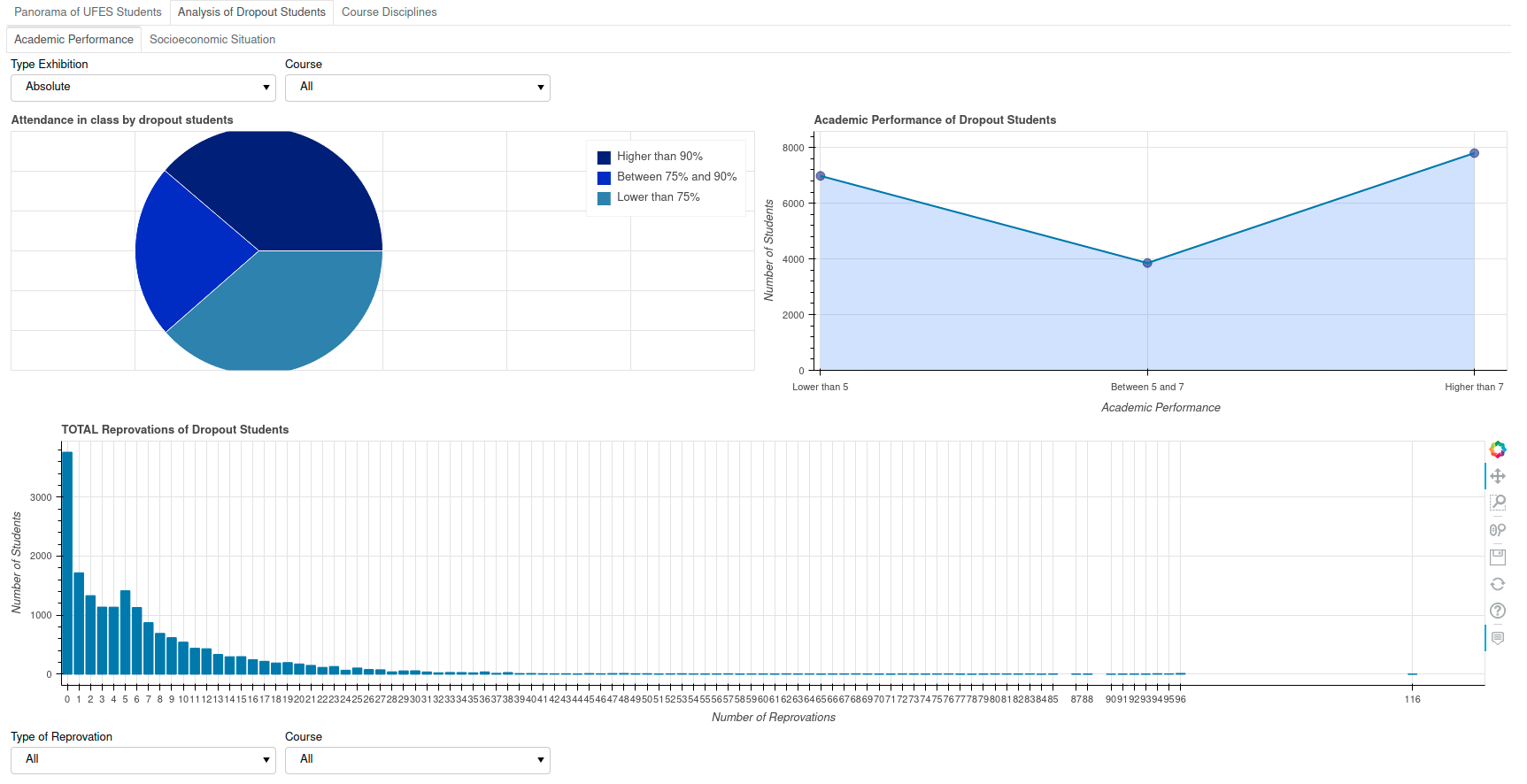}
  \caption{``Academic Performance'' sub-tab in an overview.}
    \label{figura5}
\end{sidewaysfigure}

This section presents an analysis related to students who dropped out in terms of the number of failures, percentage of absences and academic performance in the disciplines. In Figure \ref{figura5} is shown the sub-tab as a whole, in the upper left corner a pie-type graph allows to analyze the frequency of presence of students who have left from UFES; in the upper right corner there is a graph that shows three possible coefficient ranges for the student and the number of dropout students who are in each coefficient range; in the lower region there is a histogram showing the number of failures and the number of students who failed this amount of times before dropping out. There are also undergraduate course filters that cover the entire tab. 
 
This sub-tab aims to answer questions 1.e) ``What is the academic performance of the student who evades?'', 2.b) ``What is the performance rate of the highest dropout rate courses?'' and 2.c) ``What level of attendance of students who attend undergraduate courses with the highest dropout rates?'' referring to the academic performance of the student who leaves and the attendance of students of a given course. In addition, it serves as an aid to question 1.c) ``Are there courses that can observe a large or low flow of students dropping out?'' since it is possible to analyze the graphs filtered by course. 
 
\begin{figure}[h!]
  \includegraphics[width=0.9\textwidth]{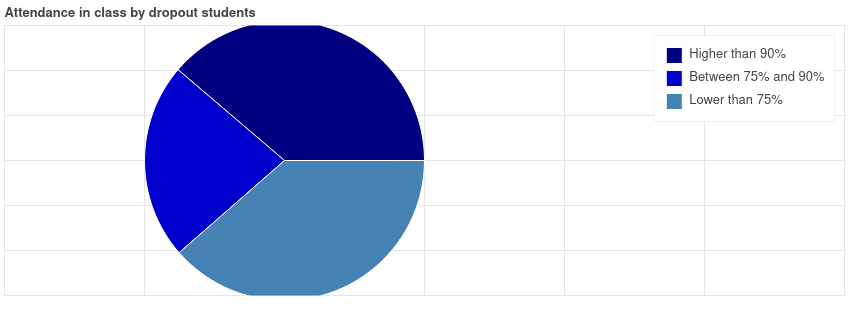}
  \caption{Graph referring to the frequency of attendance of dropout students in class. In this graph it is possible to observe the attendance distribution in the classes of dropout students. The attendance distributions chosen were greater than 90 \% of frequency in the classroom, between 90 \% and 75 \% and less than 75 \%}
    \label{figura6}

\end{figure}

The graph shown in Figure \ref{figura6} was divided into three ranges: presence above 90\%, indicates assiduous students; presence between 75\% and 90\% indicates students with a regular presence; and presence below 75\%, for students who have a level of presence that leads to failure in the Failure by Frequency modality. This index is the average attendance in all subjects that a dropout student took.
 
\begin{figure}[!h]
  \includegraphics[width=0.9\textwidth]{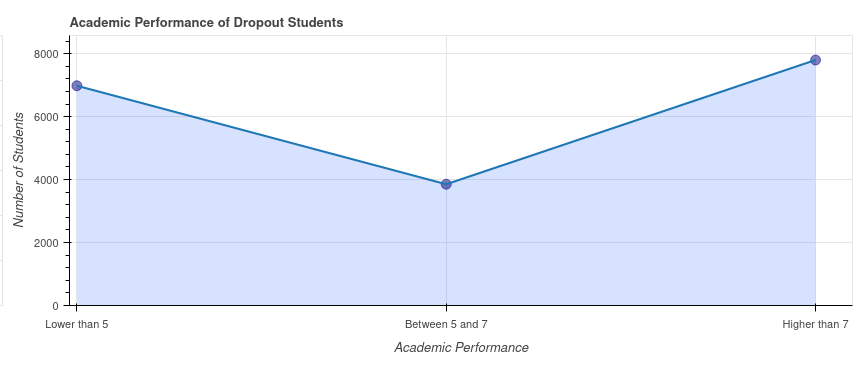}
  \caption{Graph referring to the academic performance coefficient (CR). It is possible to observe the quantitative distribution of students in each of the three determined academic performance coefficient ranges: less than 5, between 5 and 7 and greater than 7. The academic performance varies between 0 and 10.}
    \label{figura7}
\end{figure}

The graph shown in Figure \ref{figura7} deals with the performance coefficient (CR), or the weighted average of the grades taking into account the weight of the credits of each discipline. The defined ranges were the CR less than 5, for students who have an average lower than the minimum grade for obtain success in a discipline; CR greater than or equal to 5 and less than 7, for students with an average higher than the minimum grade for approval, but lower than the grade for approval without final exam; and finally an average greater than or equal to 7 for students whose average is higher than the grade for direct approval in the subjects. 
 
\begin{sidewaysfigure}[p]
  \includegraphics[width=1.0\textwidth,height=0.9\textheight,keepaspectratio]{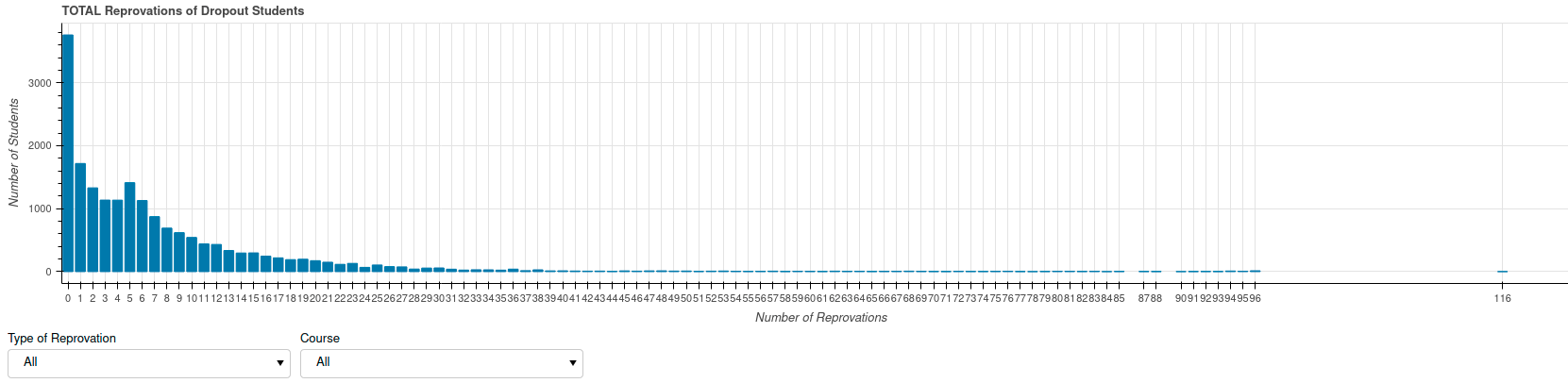}
  \caption{Histogram referring to the number of failures. In abscissa, the number of failures and in the ordained, the number of students who failed.}
    \label{figura8}
\end{sidewaysfigure}

In Figure \ref{figura8} is shown the histogram in the lower part of the sub-tab indicating the relationship between the number of failures and the number of dropout students. In this graph, it is possible to observe the tendency for students to drop out of UFES. There is also an option to choose the type of failure of the histogram where the ``ALL'' option involves the sum of the failures of the students, and the options for Grade and Frequency filter the failures of this particular type. 
 
 \subsubsection*{Socioeconomic Situation}
\begin{sidewaysfigure}[p]
 \includegraphics[width=1.0\textwidth,height=0.9\textheight,keepaspectratio]{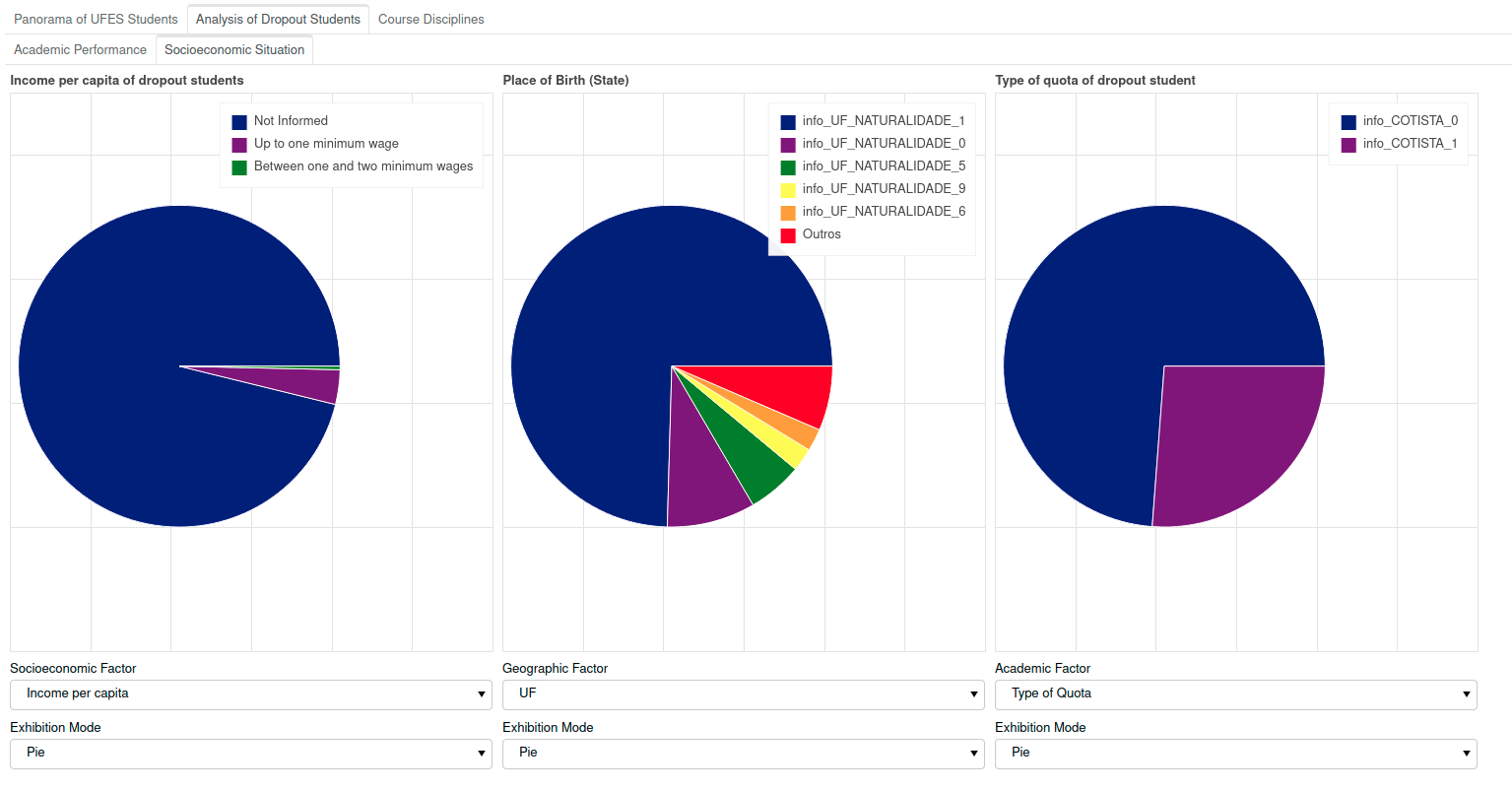}
  \caption{Graphs of the ``Socioeconomic Situation'' tab in pie mode. From left to right, graph with filters of socioeconomic factors, graph with filters of geographical factors and graph with filters of academic factors.}
   \label{figura9}
\end{sidewaysfigure}

\begin{sidewaysfigure}[p]
 \includegraphics[width=1.0\textwidth,height=0.9\textheight,keepaspectratio]{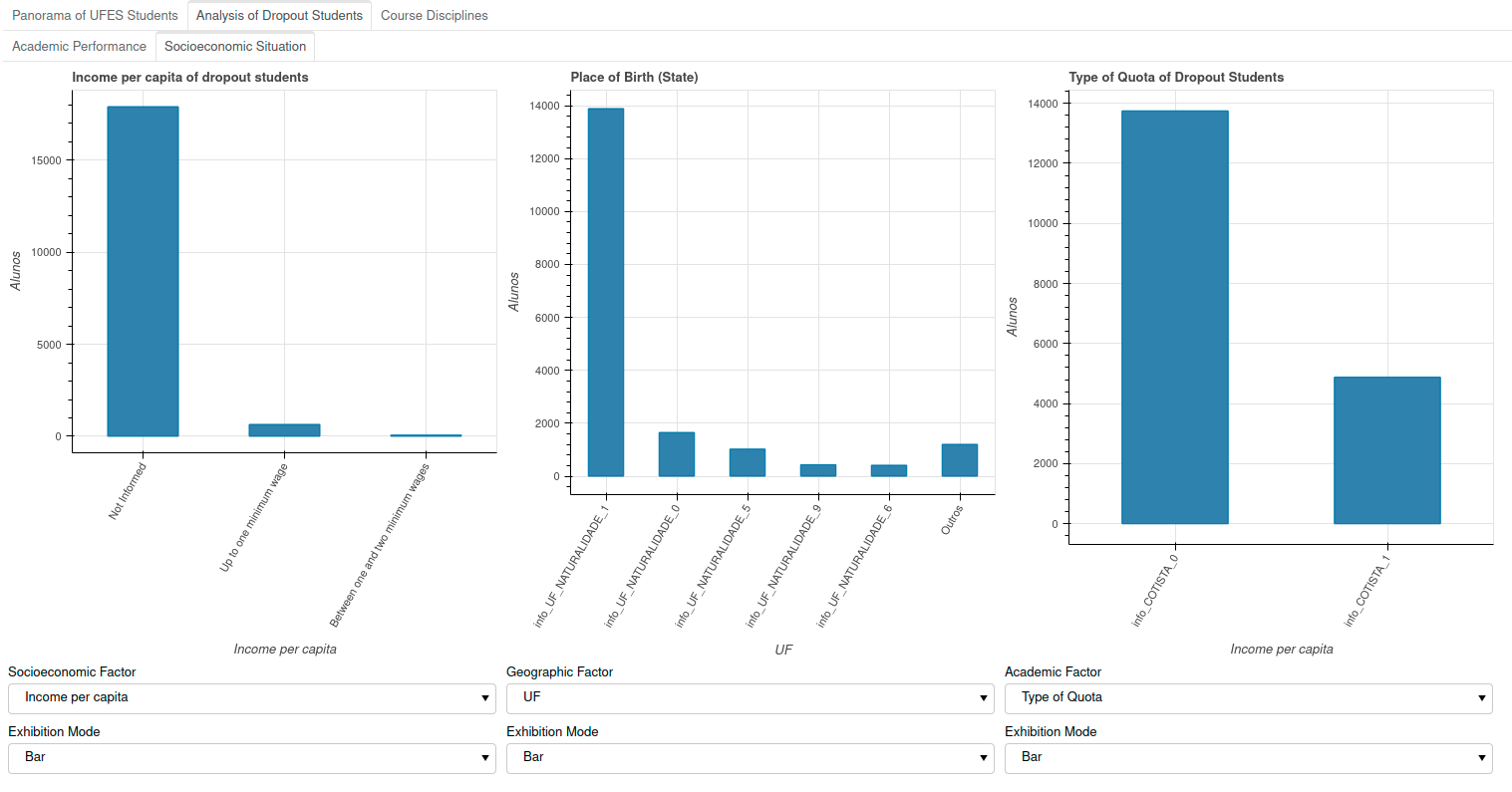}
  \caption{Graphs of the ``Socioeconomic Situation'' tab in Bars mode. Same configuration as in Figure \ref{figura9}, changing only the display mode.}
  \label{figura10}

\end{sidewaysfigure}

In this section there are three graphs, as shown in Figure \ref{figura9}, initially in pie mode, but they can be changed individually for viewing in bar mode as deployed in Figure \ref{figura10}. The first graph, from left to right orientation, indicates economic index data Income \textit{per capita}, University Aid, Housing Situation and Employment Situation; the second indicates geographic data: Place of Birth, State of Naturality and Nationality; and the third data related to the academic-social history: Type of quota, Study plan (support for students with many failures), Type of high school institution and form of admission.  
 
The graphs in this sub-tab show the five categories that had the highest number of occurrences in each of the items. In case of more than five possible categories, they are combined into a category called Others. This section aims to answer questions 1.a) ``Does the student's income influence his/her dropout?'', 1.b) ``Does the region in which the student lives affect his/her dropout?'' and 1.d) ``What is the geographic origin of the student who evades?'' referring to income and location and geographic origin of the student in order to trace the profile of the student who evades.

\subsubsection*{Disciplines}
\label{subsec-disccurso}
\begin{sidewaysfigure}[p]
  \includegraphics[width=1.0\textwidth,height=0.9\textheight,keepaspectratio]{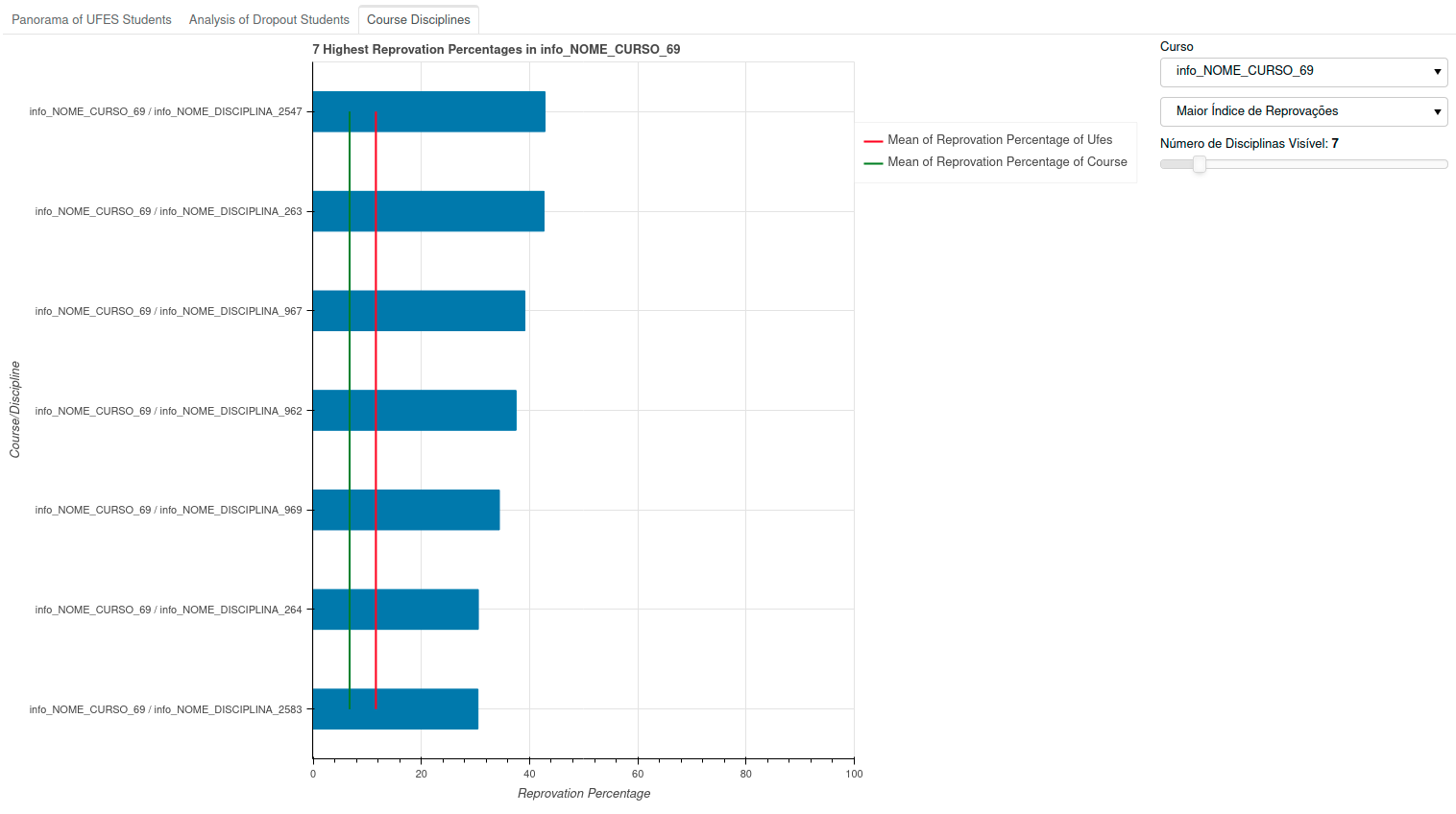}
  \caption{Graph referring highest and lowest failure rates of all disciplines of UFES. The lines represent the failing averages of UFES and the course in question for comparison purposes.}
   \label{figura11}
\end{sidewaysfigure}

This tab has only one graph, shown in Figure \ref{figura11}, which maps the subjects with the highest and lowest average failure rates. In the identifier of the discipline there is also the course to which this discipline is related, because in some situations the same discipline can be offered in different courses.
 
It is possible in this graph to filter the number of subjects shown in the graph using the slider that goes from 5 to 20 subjects, the course or, in a general case, UFES as a whole and if the subjects presented are those with a higher or lower failure rate. As a reference there are two parameters of comparison, the average of failures at UFES and in the selected course represented by the colors red and green, respectively.
 
In order to disregard cases of exception or very particular incidents and to avoid distortions, no discipline with less than 15 students enrolled was considered for this analysis. This chart serves as a tool to relate a course that has many subjects with a high failure rate to the dropout rates of that same course. Another possible relationship is a course in which the lowest failure rates are still relevant percentages. 
 
\section{Results}
\label{sec:Results}
Aiming to answer the questions proposed through graphics that can be interpreted by users, this section will clearly explain the function of each graph and how they can be used to answer the questions. 
 
\subsection*{Profile of the student who dropout from UFES}

The first question is whether there is a relationship between the student's \textit{per capita} income and dropout. This question can be answered by graphs illustrated in Figure \ref{figura2} and by the leftmost graph in Figure \ref{figura9}. 
    
Using the graph in Figure \ref{figura2}, it is necessary to apply the filter called Income \textit{per capita} and choose the income of interest to know the number of students in each situation (Graduated, No Dropout or Dropout) that fits within this income range. Thus, it is possible to understand, contextualizing for a given income \textit{per capita}, if there is an upward trend of dropout, comparing with the total number of students, with those who are still studying and those who have already graduated. 
    
The analysis of the graph shown in Figure \ref{figura9} to the left is restricted to students who have dropped out. By choosing in the filter Socioeconomic Index the option Income \textit{per capita}, it is possible to understand the fraction of the dropout students in each income range. In a previous analysis of the questions, it was defined that the analysis of the Income \textit{per capita} is extremely important in the study of student dropout, however the data referring to this factor are extremely outdated, since most students do not inform the income. Thus, it is expected that data collection will be encouraged so that the panel can contemplate this modality in the future. In addition, the Socioeconomic Index filter also offers options that allow exposing fractions of students who receive aid from the university, who have a specific housing situation and employment situation, which are subjective indexes, but can also assist in the analysis of income and in their social context.

The next question asks whether there are courses in which a large or low flow of dropout students can be observed. The graphs illustrated in Figure \ref{figura2} and Figure \ref{figura3} provide an overview of the number of dropout students and which courses have the highest dropout rates at UFES. In case of the graph illustrated in Figure \ref{figura2}, it is possible to make a course-by-course analysis, using the Course filter to compare the number of dropouts with the total number of students in different situations. By doing this analysis and understanding the context of each course, it is possible to understand the flows of dropout students and detect which courses suffer the most from this problem. 

In addition to the graphs already mentioned, there is also the graph illustrated in Figure \ref{figura4}, which indicates classes entering in 2010 and their respective progress until 2016. The student dropout measurement index in this case is the TDA at the end of the study. Courses with the highest TDA have the highest dropout flows, and those with the lowest TDA have the lowest dropout flows. 

Regarding the geographic origin of the student who dropout from higher education, as shown in Figure \ref{figura9} the middle graph refers to geographic indexes. In the Geographic Factor filter, the options State, Place of Birth and Nationality appear. The State option refers to the state or province of origin of the students, the Naturalness option refers to the city of origin of the students and the Nationality option refers to the country of origin of the students. Through this graph it is possible to observe each fraction corresponding to the city, state and country of origin of the students. 
	 
A barrier found in the creation of the graphs illustrated in Figure \ref{figura9} was the possibility of reducing the categories represented in the graphs. This functionality could hide some category of great predominance to understand how the other categories are related, for example, in the case of students from outside Espírito Santo or Brazil.

As academic performance, raised in the question 1.e) ``What is the academic performance of the student who evades'' related to the profile of students who drop out, one can understand several aspects performed by a student, ranging from the number of failures through attendance and also by the CR. There are several graphs on the panel that analyze aspects of the student's academic performance. There is also a dedicated tab for these aspects shown in Figure \ref{figura5}. The graph illustrated in Figure \ref{figura6} analyzes the students' attendance, which is covered in the Section on Dropout in the Institution. 
	
In Figure \ref{figura7}, a line graph showing three CR ranges related to the number of dropout students from UFES can be observed. The chosen CR ranges were related to the UFES approval regulations. At the end of the period, if the student scores above 7 he/she is approved, if below 7 he/she is entitled to a final exam. If the arithmetic average between the partial grade and the final exam score is greater than 5, then the student is approved. The three bands were then defined: below 5, between 5 and 7 and above 7. There are two filters referring to this graph, one that chooses the way of displaying the data, being able to choose percentage or absolute value, and another that the course of which the data are referring. This graph provides an overview by course of what is the tendency of the grade of students who drop out.
	
The graph shown in Figure \ref{figura8} is a histogram that relates the frequency of students to the number of failures. In other words, it is possible to observe in the graph the number of dropout students who did not fail, who failed once, twice and so on. In the graph it is possible to filter the type of failures in three options: failures by grade, by frequency and all failures. It is also possible to choose the course of interest for the analysis of failures. Through this graph you can have an overview of the trends of high numbers of failures by the course and by the type of failure.

\subsection*{Dropout in the Institution}

The first question related to dropout at the institution seeks to identify the courses with the highest dropout rates at UFES. The graph illustrated by Figure \ref{figura3} indicates a ranking with the 15 courses with more or less students in a given situation. When choosing the Dropout situation, it is possible to observe a ranking of the 15 courses with the highest or lowest number of dropout students. This could be an answer to this question, but according to UFES data, it is not possible to identify courses by name, which implies in a lack of context for analysis. It is not possible to compare, for example, a course that has only one class formed with a course that has existed at UFES for decades. To solve this problem, the course of interest can be analyzed in the graph illustrated in Figure \ref{figura2}, in order to highlight the number of students dropping out in perspective with the total number. 

Next, it is asked what is the performance rate of these courses. As explained earlier, the graphs presented in Figure \ref{figura5} analyze certain aspects of academic performance and have course filters. Allied to the graphs illustrated in Figure \ref{figura1}, where it is possible to define the courses with the highest dropout rate, such courses of interest to the study can be used. In addition to this relation, there is a graph that assists in this analysis, which is shown in Figure \ref{figura11}, it is a ranking of disciplines that have a higher or lower failure rate and their corresponding courses. Through this graph it is possible to expose the disciplines that are most challenging in a given course, and in some cases, to detect disciplines whose failure rates are very high. Frustration with the difficulty of the course can be a discouraging factor, which can lead to dropout \citep{alyahyanpredicting}. 

Regarding the level of attendance of students who attend undergraduate courses with the highest dropout rates, the graph in Figure \ref{figura6} in a pie format indicates the fraction of the number of dropout students at different levels of attendance along their trajectories at UFES. The defined ranges were in percentage of presence in class for very assiduous students, students with regular attendance within the normal range and students who have an insufficient percentage of presence in class for approval. Using the Course filter, above the graph, it is possible to define the analysis course. The question refers to the courses with the highest dropout rates at the university. Although the choice in the Courses filter of the pie chart illustrated in Figure \ref{figura6} is a manual step for the user, there are also other charts that expose the courses with the highest dropout rates. The graphs illustrated in Figure \ref{figura1} provide an overview of the number of dropout students in each course and which courses have the highest number of students in situations of Dropout at UFES. In addition, in the graph illustrated in Figure \ref{figura4}, it is possible to observe courses that have high TDA. However, as the UFES data is anonymized, it is not possible to draw a parallel regarding the TDA and the attendance given in the pie chart in Figure \ref{figura6}. 

\section*{Conclusion}
\label{sec:Conclusion}
The development of this study aimed at creating a data visualization tool to facilitate the exploratory analysis of aspects relevant to university dropout in Higher Education Institutions. The tool is made of a dashboard that is the materialization of the exploratory analysis of student dropout of data collected by UFES and INEP. The panel also supports an update of the data, as long as its main characteristics of column quantities and their nomenclatures are maintained, so that the analysis progresses over time. It is important to note that the developed panel is an auxiliary tool that loses its meaning without a well-defined context of the HEI studied. This study defines a methodological model for the development of similar tools in other HEIs, taking into account the different datasets and possible different metrics that can enrich the analysis.

Through the dashboard developed, it was possible to outline the profile of the student who evades. Eight questions were proposed, as described in the Section~\nameref{sec:Intro}, aiming to profile the general characteristics of a dropout student and the institution from which he/she dropped out. This information, derived from such answers, is related to social, geographic, income, academic performance and attendance factors of the student and how the disciplines and courses are positioned in relation to dropout and failure in a spectrum of exploratory data analysis. The answer to these questions, through graphics, served as an indicator of the profile of the student and the institution, which could be properly exposed on the panel. It was not always possible to find a pattern in the profile of the student who evades. At times, due to some gaps in the data (the overwhelming majority of students do not report their family income \textit{per capita}, for example) it was difficult to draw profiles in some of the categories. In other cases, it was possible to use the information even to clarify that a certain factor is not directly related to dropout. However, through the panel's data and assistance, it is possible to obtain answers and define whether or not there is a pattern among students.

The proposed questions could be answered in whole or in part by the graphs presented on the dashboard. Most of the questions related to UFES could be answered in a generic way, as the data are anonymized. The only question that has an indeterminate answer is ``Does the student's income influence their dropout rate?'', in which most of the enrolled students do not inform their income. However, as the graph used to reach the conclusion that most student's income information is not informed, a reduction of missing data in the dataset would be enough to obtain more accurate answers through the panel. 

 Various data were obtained with a coded name, and therefore, it was hard to carry out an analysis. The panel allows to answer the proposed questions, but an analysis, for example, of the course that has more dropout students in the dataset provided by UFES, would be incomplete. It is possible to see the percentage of dropout students represented in the historical series of this course (67\%), but the analysis is only properly done when considering the whole set of name, history and context of such course at UFES. 

On the other hand, INEP data are quite explicit and show a worrying reality. In a group of classes that started in 2010 and analyzed until 2016, the calculated TDA averages, both national, such as from the state Espirito Santo and even those from UFES, are very high. The lowest among them is precisely that of UFES, with 44.1\%. The national average is 56.5\%. Analyzing the data from UFES, it was possible to see that the courses had completion time for a maximum of 2016. That is, by the end of 2016, all of these analyzed classes should be in the process of completing the course or already completed. Even with an accumulated dropout average lower than the national average, UFES indexes are close to half of the incoming students. This reinforces the importance of the attention that should be directed to this topic. With such high dropout rates, we see financial and human resources for education being wasted. Brazilian public HEIs produce a great part of the scientific knowledge of the country according to a report by Clarivate Analytics, 15 of them are responsible for 60\% of Brazilian scientific production \citep{analytics2018research}, but the scenario could be even better if the dropout rates were not so high.

This work was done in order to answer general questions regarding the profile of the student and the HEIs. Improvements are needed in order to define more specific questions and answer them in a more elaborate way. Another important aspect is the quality of the data provided, where some information, such as income \textit{per capita}, were left with almost no information, and many information from some students came repeated. Another interesting aspect to improve data quality would be the inclusion of the distance between the student's home to the HEI, as this was one of the questions raised. The problem of the distance between the student's residence in relation to the university can be a relevant factor in dropout. An improvement in the quality of data required by UFES for new entrants will not impact the analysis of classes already formed, but is encouraged to improve future analyzes.

Some limitations of the Bokeh library, which could be overcome by incorporating parts of code from other languages, such as javascript and HTML \citep{manualbokeh}, prevented the application of certain tools, such as the selection of categories presented in the graphics in the pie format in Figure \ref{figura9}. This type of incorporation was lacking in some items of the panel that could be more complete and produce more effective analyzes. A suggested improvement is to use web languages built into the dashboard to improve data visualization. Another interesting point for the panel would be improve the use of web content development guidelines that allow accessibility \citep{caldwell2008web}. 

It is worth to mention that given the current global pandemic circumstances, distance measures and changes in the paradigm related to how the HEIs work, the impact on student dropout rates is unknown. This promotes the importance of the panel being used with updated data so that, in the future, we can understand the effect that the pandemic had on the dropout flows of the HEIs. Several studies are already being conducted regarding the impacts of the Covid-19 pandemic on the mental health and motivation of university students \citep{meetercollege, arslanloneliness, coughenourchanges}. Such psychological factors can directly impact the university dropout rates \citep{alyahyanpredicting} and, although the psychological aspect has not been addressed in the panel, they can be incorporated into a possible continuity of the panel if data are collected for this purpose. 

The continuity of the development of the panel is necessary for improvements to be made, and a response from the public and managers is essential to feed the progress of the tool. There may be an analysis of new research related to student dropout and the discussion among stakeholders to discover new relevant questions so that they can be answered through the visualization on the panel.

\bibliography{main}

\end{document}